\documentclass[twocolumn]{revtex4}

\usepackage{graphicx} 

\begin{document}

\title{Status of the Milagro Gamma Ray Observatory}

\newcommand{\uw}{$^{1}$}
\newcommand{\ucsc}{$^{2}$}
\newcommand{\um}{$^{3}$}
\newcommand{\gmu}{$^{4}$}
\newcommand{\unh}{$^{5}$}
\newcommand{\nyu}{$^{6}$}
\newcommand{\lanl}{$^{7}$}
\newcommand{\uci}{$^{8}$}
\newcommand{\ucr}{$^{9}$}

\newcommand{\adduw}[1]{$^{1}$ #1 }
\newcommand{\adducsc}[1]{$^{2}$ #1 }
\newcommand{\addum}[1]{$^{3}$ #1 }
\newcommand{\addgmu}[1]{$^{4}$ #1 }
\newcommand{\addunh}[1]{$^{5}$ #1 }
\newcommand{\addnyu}[1]{$^{6}$ #1 }
\newcommand{\addlanl}[1]{$^{7}$ #1 }
\newcommand{\adduci}[1]{$^{8}$ #1 }
\newcommand{\adducr}[1]{$^{9}$ #1 }

\author{
{\large The Milagro Collaboration} \\ 
\bigskip
R.~Atkins\uw,
W.~Benbow\ucsc,
D.~Berley\um,  
E.~Blaufuss\um,
J.~Bussons\um,
D.~Coyne\ucsc,
S.~Delay\uci,
T.~DeYoung\ucsc,
B.~Dingus\uw,
D.~Dorfan\ucsc,
R.~Ellsworth\gmu,
A.~Falcone\unh,
L.~Fleysher\nyu,
R.~Fleysher\nyu,
G.~Gisler\lanl,
J.~Goodman\um,
T.~Haines\lanl,
E.~Hays\um,
C.~Hoffman\lanl,
S.~Hugenberger\uci,
L.~Kelley\ucsc,
I.~Leonor\uci,
J.~McCullough\ucsc,
J.~McEnery\uw,
R.~Miller\lanl,
A.~Mincer\nyu,
M.~Morales\ucsc,
P.~Nemethy\nyu,
D.~Noyes\um,
J.~Ryan\unh,
F.~Samuelson\lanl,
M.~Schneider\ucsc,
B.~Shen\ucr,
A.~Shoup\uci,
G.~Sinnis\lanl,
A.~Smith\um,
G.~Sullivan\um,
T.~Tumer\ucr,
K.~Wang\ucr,
M.~Wascko\ucr,
S.~Westerhoff\ucsc,
D.~Williams\ucsc,
G.~Yodh\uci \\
\smallskip
\footnotesize
\it
\adduw{Department of Physics, University of Wisconsin, Madison, WI 53706}\\
\adducsc{Department of Physics, University of California, Santa Cruz, CA 95064
}\\ 
\addum{Department of Physics, University of Maryland, College Park, MD 20742}\\
\addgmu{Department of Physics, George Mason University, Fairfax, VA 22030}\\
\addunh{Department of Physics, University of New Hampshire, Durham, NH 03824}\\
\addnyu{Department of Physics, New York University, New York, NY 10003}\\
\addlanl{Physics Division, P-23, Los Alamos National Laboratory, Los Alamos, NM 87545}\\
\adduci{Department of Physics, University of California, Irvine, CA 92717}\\
\adducr{Department of Physics, University of California, Riverside, CA 92521}\\
}
\affiliation{ }

\begin{abstract}
The Milagro Gamma Ray Observatory, located at an altitude of 8,600 feet in the 
Jemez Mountains of New Mexico, is the world's first large-area water Cherenkov 
detector capable of continuously monitoring the entire sky for sources of TeV gamma rays. It is uniquely capable of searching for transient sources of 
VHE gamma rays. The core of the detector is a 60m x 80m x 8m pond 
instrumented with 723 PMTs deployed in two layers. This part of the detector is complete and has 
operated continuously since Jan. 2000. Initial studies including searches for gamma-ray sources are 
ongoing, and preliminary results are available. The final stage of construction is 
under way. We are deploying 170 auxiliary "outrigger" water Cherenkov 
detectors in an area of 40,000 square-meters surrounding the pond, which will 
significantly enhance our ability to reject background and more accurately 
reconstruct the gamma-ray direction and energy. In addition, we are lowering the 
energy threshold of the detector by using custom processing to enable real-time 
intelligent triggering. The lower energy threshold will significantly increase our 
sensitivity to gamma-ray sources, and in particular to sources of cosmological 
origin, such as GRBs, where the higher energy gamma-rays have sizable 
attenuation due to the interaction with the intergalactic infra-red light.
\end{abstract}

\maketitle

\section{Introduction}
The observation of high-energy gamma ray sources has helped us to gain a better understanding of the energetic acceleration processes in the Universe. The known gamma ray sources include supernova remnants, active galactic nuclei (AGN), and gamma ray bursts (GRB). In addition, more exotic sources like topological defects, evaporating primordial black holes, and dark matter particle annihilation and decay could be sources of high-energy gamma ray emission. 
Several reviews 
of the techniques, science, and recent results in high-energy gamma-ray 
astronomy have been published~\cite{Hoffman,Ong,Weekes}.

The Milagro detector is a new type of instrument designed to search for continuous and episodic sources of VHE gamma ray sources. 
Milagro is the first detector sensitive to cosmic gamma rays below 1 TeV with the all-sky, high duty-factor capabilities of an extended air shower (EAS) array. Our prototype detector, 
Milagrito, has produced results on the outburst of the AGN Mrk501 and given suggestive evidence for TeV emission associated 
with a GRB~\cite{grito,mrk501,grb97}. We describe here the Milagro detector and its performance, as well as the current status of operations. This paper was presented at the 2001 ICRC~\cite{sullivan}. For details on the various physics topics addressed by Milagro see other talks in those proceedings~\cite{Benbow,Fleysher,Ryan,Samuelson,Sinnis,Smith,Stephens,Yodh}. 
%
 \begin{figure}[t]
 \vspace*{0mm} 
 \includegraphics[width=8.3cm,height=4.0cm]{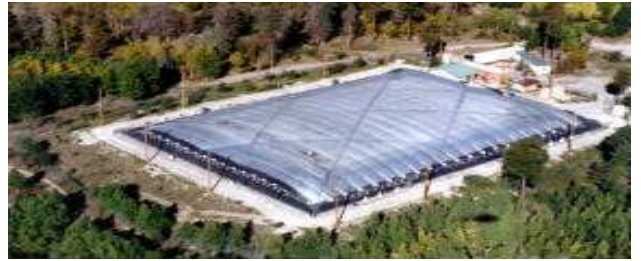} 
 \caption{Aerial view of the Milagro pond with the light-tight cover inflated for work inside..}
 \end{figure}

\subsection{Scientific Goals of Milagro}
Milagro performs high-duty-factor, all-sky observations in the VHE region. 
We are using the Milagro detector to: 
\begin{itemize}
\item Survey the Northern sky for steady and episodic 
sources.
\item Search for emission from GRBs in the energy range ~100 GeV to 20 TeV.
\item Detect VHE emission from the Crab and study its energy spectrum with 
a new, independent technique. 
\item Continuously monitor the entire sky for  
flaring sources such as Markarian 421 and Markarian 501. 
\item Search for emission from the galactic plane. 
\item Detect the shadow of the moon with high significance and use it to 
calibrate the energy response of the detector and to perform a search for TeV cosmic antiprotons. 
\item Detect the shadow of the sun with high significance and use it to 
continuously monitor the strength of the transverse component of the solar 
magnetic field. 
\item Search for evaporating primordial black holes.
\item Search for $>$5 GeV particles from the Sun with unprecedented sensitivity.
\item Measure the primary proton spectrum with single hadrons
\item Measure the composition of the cosmic rays above 50 TeV in conjunction with WACT
\item Search for WIMPS annihilating in the vicinity of the Sun.
\end{itemize}

\section{The Milagro Detector}
Milagro uses photomultiplier 
tubes (PMTs) deployed under water to detect the 
Cherenkov 
radiation produced in the water by relativistic charged shower particles. 
Because water is inexpensive and the Cherenkov cone spreads out the light, one 
is able to construct a large instrument that can detect nearly every charged 
shower particle falling within its area. Furthermore, the plentiful photons 
convert to electron-positron pairs (or to electrons via Compton scattering). 
These electrons, in turn, produce Cherenkov radiation that can be detected. 
Consequently, Milagro has an unprecedented low energy threshold for an 
EAS-array.
%
 \begin{figure}[t]
 \vspace*{0mm} 
 \includegraphics[width=8.3cm,height=5.0cm]{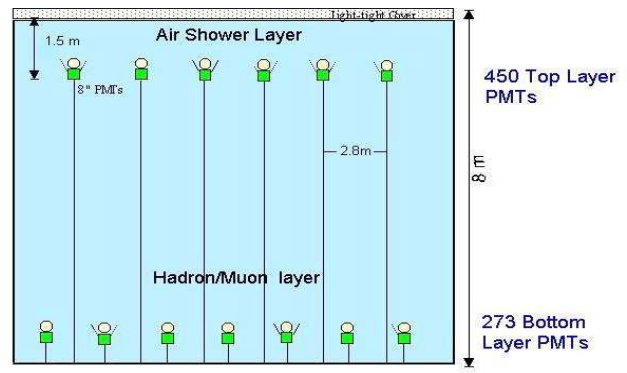} 
 \caption{Schematic cross section of the Milagro detector.}
 \end{figure}
As in a conventional EAS-array, the direction of the primary gamma ray is reconstructed in Milagro by measuring the relative arrival time of the shower front across the detector. 

The Milagro detector is located at an altitude of 8,600 feet (750 $g/cm^{2}$ ) at the Fenton Hill site in the Jemez Mountains of New Mexico (Figure 1). 
The core of the detector is a 6-million gallon pond measuring 60m x 80m x 8m (depth), which is used as a large area water Cherenkov detector. 
When completed the pond will be surrounded by 170 individual water Cherenkov 
detectors, called ``outriggers'', over a 200m x 200m area.

The Milagro pond is covered by a light-tight barrier and instrumented with an array of 450 photomultiplier tubes 
deployed under 1.5-m of water to detect air-shower particles reaching the 
ground. These PMTs measure the arrival time and density of the air-shower 
particles. In addition, 273 PMTs are located at the bottom of the pond 
under 6m of water and are used to distinguish photon-induced showers 
from hadron-induced showers. The top array of PMTs is called the shower 
layer and the bottom array is the muon layer (Figure 2). On the bottom of the pond is a 2.8m x 2.8m grid of sand-filled PVC pipe to which The PMTs in both layers are secured by a Kevlar string, as shown in Figure 2.
Milagro uses 20-cm-diameter 
Hamamatsu 10-stage R5912SEL PMTs. Custom-made front-end electronics boards provide timing and charge 
information for each PMT channel. 
The front-end boards also provide triggering information. 

An event display for a typical air shower event in the Milagro pond is shown in Figure 3.
The line above the pond indicates the best fit direction of the shower front using the relative arrival times of the shower layer PMTs. 
%
 \begin{figure}[h]
 \vspace*{0mm} 
 \includegraphics[width=8.3cm,height=7.0cm]{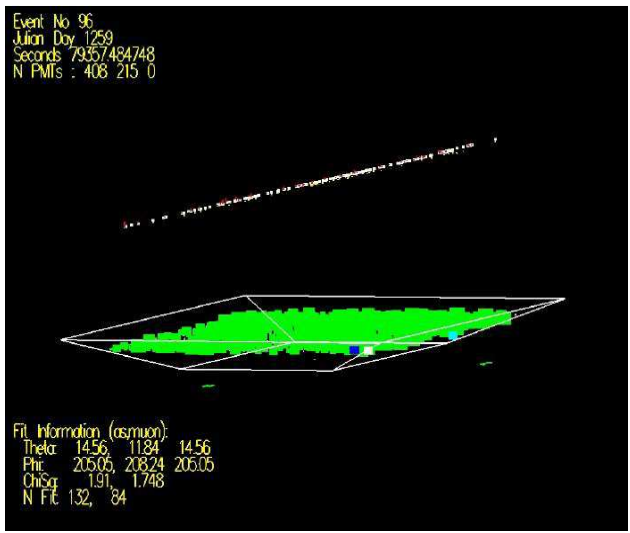} 
 \caption{Air Shower event in Milagro. The line above the pond is the timing fits from the Shower  layer. The green squares are proportional in size to the signal in the Shower layer.}
 \end{figure}

A measure for the angular resolution of the Milagro detector is the  DELEO/2  distribution of Figure 4. DELEO/2 is one half the space angle difference between the fits using the  odd  and  even  tubes in the array and is approximately the angular uncertainty (excluding certain systematics). The final resolution of the detector will be degraded by uncertainties in the shower's core position coupled with the shower-front's small curvature. The angular resolution is dependent on the number of PMTs that are in the fit; the resulting average angular resolution for all reconstructed events is $\sim 0.75^{\circ}$.
%
\begin{figure}[h]
\vspace*{0.0mm} 
\includegraphics[width=8.3cm,height=4.5cm]{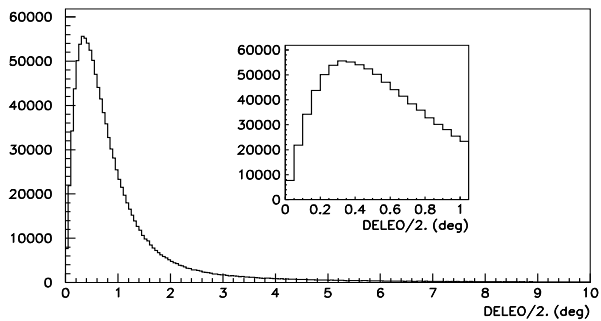} 
\caption{Distribution of "DELEO/2" for Milagro.}
\end{figure}

Figure 5 shows the effective area of the detector versus gamma ray energy for the current trigger (blue) and planned trigger discussed below (black). The Milagro detector becomes sensitive to gamma-ray induced showers of an energy of $\sim$200 GeV. For a gamma-ray energy spectrum of $E^{-2.4}$ the median energy for all triggers, events reconstructed within a $2.1^{\circ}$ bin , and events that pass the gamma-hadron cut are 6 TeV, 3 TeV, and 4.7 TeV respectively. 
%
\begin{figure}[h]
\vspace*{0.0mm} 
\includegraphics[width=8.3cm,height=6.0cm]{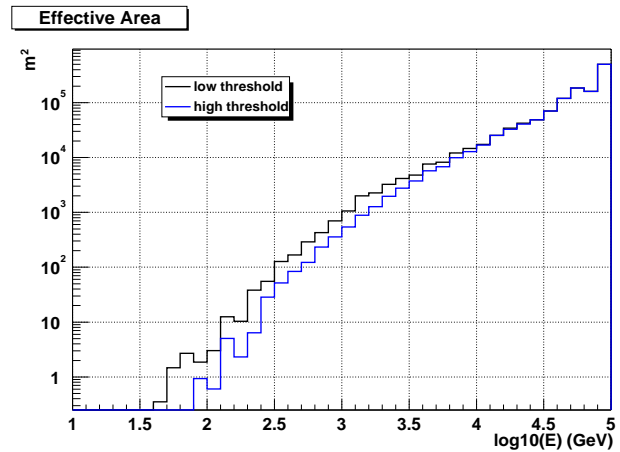} 
\caption{Milagro's effective area to gamma-ray showers versus energy. }
\end{figure}

\section{Current Operations and Results}
The construction of the Milagro pond was completed in early 1999, and operations began in mid 1999. After a three month shut down in the fall, the detector has operated nearly continuously since November 1999. Since Jan. 2000 the detector has operated at a trigger rate of 1,500-2,000 Hz, depending on experimental and meteorological conditions (Figure 6). All events are reconstructed in real time at the site. The entire data set of reconstructed events is copied over the network to our archival data storage disk array
and stored in a highly compressed format. It takes approximately 1.3 Terabytes of disk to store one year of all-sky data. In addition to the all-sky data set, there are ``source'' files that contain the full data for selected sources and the sun and moon. These files are recorded on DLT tape.
%
 \begin{figure}[h]
 \vspace*{0mm} 
 \includegraphics[width=8.0cm,height=9.0cm]{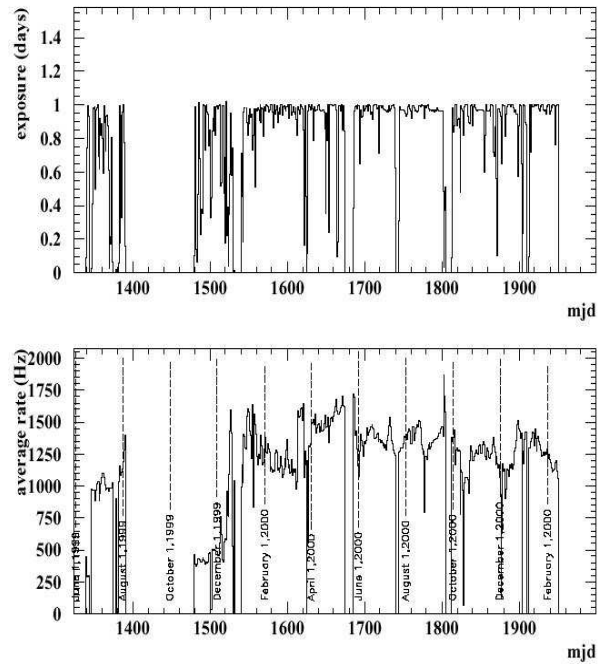} 
 \caption{Exposure of the Milagro detector since turn on in early 1999.}
 \end{figure}

In Milagro, a gamma-ray signal from a source appears as an excess of events from 
the source direction, compared with the background from hadronic 
cosmic-ray showers. An important feature of the Milagro detector is its ability to reduce this hadronic background by using the muon layer in the pond. This gamma-hadron 
separation is described in detail elsewhere in these proceedings 
(Sinnis, 2001). Figure 7 shows the Crab signal accumulated versus time, and figure 8 
shows the significance of the event excess in the vicinity of the Crab during the period June 8, 1999 to April 24, 2001. At the source 
position, an excess of 4443 events  
is observed, corresponding to a significance of 4.8$\sigma$. 
%
 \begin{figure}[h]
 \vspace*{0mm} 
 \includegraphics[width=8.0cm,height=9.0cm]{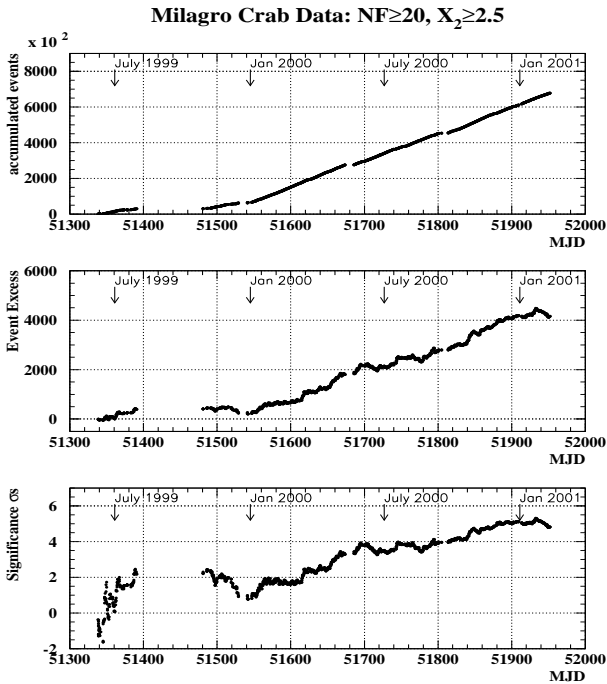} 
 \caption{Milagro's Crab signal versus time. The plots show the accumulated events(top), event excess (middle), and signal significance (bottom) at the position of the Crab nebula.}
 \end{figure}

%
 \begin{figure}[h]
 \vspace*{0mm} 
 \includegraphics[width=8.0cm,height=7.0cm]{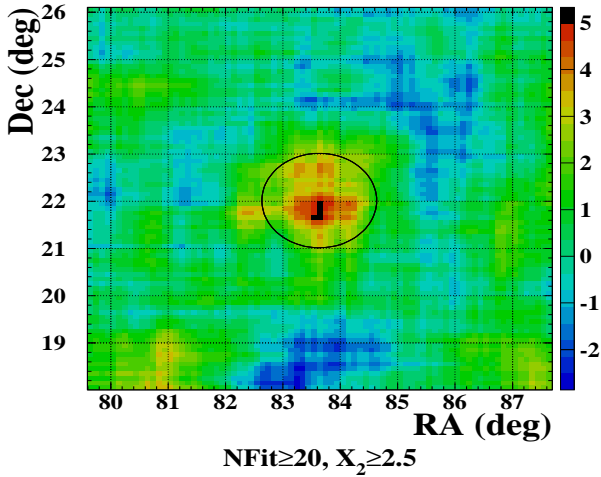} 
 \caption{Sky-map of the signal around the Crab. The colors represent the excess in sigma, with the scale at the right of the plot. The black circle is centered on the true Crab position.}
 \end{figure}

We have also observed gamma-ray emission from Mrk421, which has recently been observed to be active in both the X-ray, by the All-Sky Monitor(ASM), and at TeV energies by HEGRA and Whipple. During the period June 14, 2001 - April 24, 2001 we find an event excess corresponding to 4.0$\sigma$ (Benbow, 2001). We have also observed the shadow of the moon (Samuelson, 2001), and performed a search for GRB (Smith, 2001). 

\section{Future Improvements}
Since the construction of the Milagro pond was completed the detector has become operational and successfully observed VHE gamma-ray sources. However, there are two significant improvements that are now being implemented at Milagro, which will significantly improve its sensitivity. The first improvement is to finish the construction of the detector by deploying $\sim$170 individual water tanks as Cherenkov detectors, called ``outriggers'', surrounding the pond. The second is the implementation of smart triggering processors that will  significantly lower the energy threshold.

\subsection{Outrigger Water Tank Array}
Approximately 70\% of the showers that trigger the Milagro detector have cores that do 
not fall directly on the pond. It is vital to be able to determine the shower 
core position to substantially improve the performance of Milagro with respect 
to angular resolution, gamma-hadron separation and energy determination of 
each event. Without knowing the actual position of the core it is difficult to tell a low energy shower hitting the pond from a larger energy shower with a core far away. Additionally, because the shower front is curved, not knowing the core position can give a systematic pointing error ($\sim 1^{\circ}$) degrading the angular resolution and therefore sensitivity. Finally, our ability to perform gamma-hadron separation using the muon layer will be enhanced by an increased knowledge of the true core position.

In order to identify the core position we are deploying $\sim$170 individual cylindrical water tanks, 0.91 m in height by 2.4 m in diameter, made of polyethelene and lined with Tyvek. The tanks are instrumented with a single PMT facing down from the top, which enable them to act as individual water Cherenkov detectors to measure the particles in the EAS with high efficiency. These tanks will surround the pond in an area approximately 200m x 200m with the layout shown in Figure 9. In addition to improving our energy resolution, studies show that the outrigger array will increase our sensitivity to gamma-ray sources by at least a factor of two.
%
 \begin{figure}[h]
 \vspace*{0mm} 
 \hspace*{5.0mm}
 \includegraphics[width=8.0cm,height=8.0cm]{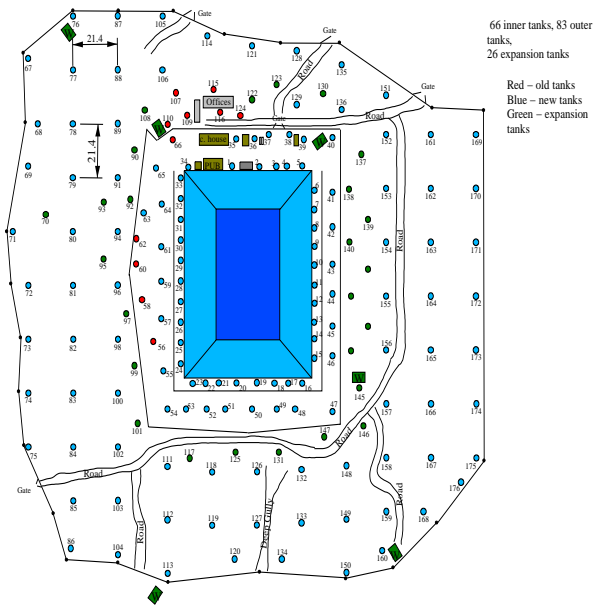} 
 \caption{Plan for the final layout of the Milagro detector. The pond will be surrounded by $\sim$170 outriggers.}
 \end{figure}

\subsection{Lowering The Energy Threshold}
The Milagro detector is capable of high reconstruction efficiency and good angular resolution for gamma-ray showers down to $\sim$10 PMTs. Currently, the low energy reach is limited by the trigger system, which is a simple multiplicity of PMTs in the shower layer. The DAQ system's readout capability is about 2kHz, and using a simple multiplicity trigger this rate is reached at $\sim$55 PMTs. This 55 PMT trigger threshold is well above the detector's intrinsic capability to detect and reconstruct gamma rays down to $\sim$10 PMTs. Below the threshold of 55 PMTs the trigger rate becomes rapidly dominated by non-shower events such as single muons, and the total rate is higher then our ability to read out. In order to reject these non-shower events, we are developing custom trigger processors using the time signature of the PMTs hit and muon layer information to reduce our threshold. We are currently implementing these custom processors, and hope to update our progress at the conference.

A lower energy threshold will significantly increase our sensitivity to gamma-ray sources, and in particular to sources of cosmological origin, such as GRBs, where the higher energy gamma-rays have sizable attenuation due to the interaction with the intergalactic infra-red light~\cite{Hays}. 

\section{Conclusion}

Milagro is the first EAS detector with an all-sky, high duty-factor capability, to be sensitive to gamma rays in the TeV energy range. The Milagro pond, which is the core of the detector, has been operational continuously since Jan. 2000 and has observed two TeV gamma-ray sources. Planned improvements will further increase our sensitivity, and put us in a unique position to monitor the sky for episodic emission of VHE gamma ray sources.

This work is supported by the US National Science Foundation, and also by the DoE, the LDRD program at LANL, and the IGPP.

\end{document}